**Translation-Rotation Coupling and the Kinematics of Non-Slip Boundary Conditions: A Rough Sphere between Two Sliding Walls**

Yueran Wang and Peter Harrowell[*]

*School of Chemistry, University of Sydney, Sydney New South Wales 2006 Australia*

* corresponding author:  peter.harrowell@sydney.edu.au

Abstract

A non-slip constraint between a particle and a wall is applied at the microscopic level of collision dynamics using the rough sphere model. We analyse the consequences of the translation-rotation coupling of the rough sphere confined between two parallel planar walls and establish that shearing the walls past each other i) preferentially deposits energy into the rotational degree of freedom and ii) results in a bounded oscillation of the energy of the confined particle.

## 1. Introduction

The dynamic boundary condition between a wall and the adjacent fluid represents the essential constraint that characterises a hydrodynamic flow. At the macroscopic level, the boundary conditions at a wall are often characterised by a slip length [1], a formalism that allows for a continuous adjustment of the boundary condition from slip (i.e. an infinite slip length) to stick (with a slip length zero). The stick boundary condition corresponds to a constraint on the velocity of the fluid in contact with the wall, namely

$$v_t^{wall} = V_{wall} \qquad\qquad (1)$$

where $v_t^{wall}$ is the transverse component of the particle velocity adjacent to the wall. At a microscopic level, the imposition of Eq.1 is problematic as it would require a discontinuous change in the magnitude of particle velocities on contact with the surface. In reality, the dynamics of particles adjacent to a wall are determined by the morphology of the wall surface and the attractive interactions between wall and particle [2]. In this picture, stick boundaries arise as an approximate expression of the impediment to transverse motion imposed by the microscopic variations in wall surface. Such microscopic considerations lead directly to boundary conditions that are more complex than the simple stick condition expressed in Eq.1 and dependent on the details of wall roughness and liquid particle interactions. This complexity of the liquid-wall coupling has given rise to a considerable literature on shear rate dependent slip at the liquid-wall interface [3-5]. To establish the generic consequences of the microscopic coupling between particle and wall, it would be useful to have a minimal model that is simple enough to analyse in detail and generic enough for the results of this analysis to be of general use.

In this paper we shall study the dynamics of a rough sphere confined between sliding parallel walls. The rough sphere, as described below, represents a simple expression of a microscopic non-slip condition. Of particular interest is the role of the coupling between translational and rotational motion that arises as an unavoidable consequence of the non-slip condition applied to the rough sphere – wall collisions.

The distinguishing feature of the rough sphere is that on collision, the transverse component of the velocity of surface of the sphere relative to the wall is reversed. This is in addition to the reversal of the normal velocity that is standard for elastic collisions of smooth particles. This simple model is of sufficient interest and utility to have been independently invented at least three times, each in a different physical context. It was first proposed by Bryan in 1894 [6]. In 1922 Pidduck [7] followed up Bryan's idea and derived the Chapman-Enskog equations for the kinetics of a dilute glass of rough spheres. Dahler and coworkers [8,9] further developed the kinetic theory of rough sphere gases in 1966. Chapman and Cowling included an extensive discussion of the model in the 1970 edition of their book on gas kinetic theory [10]. In 1963 Waldmann [11] considered the kinetics of the rough sphere Lorentz gas and noted that translation-rotation coupling results in a significant decrease in the diffusion constant relative to that of the hard sphere. The model was extended to dense liquids by O'Dell and Berne [12] and Kravchenko and Thachuk [13,14] . The rough sphere's second incarnation came in late 1960's, when Strobel [15] and Garwin [16] rederived the kinematic equations for the model, inspired, in part, by the appearance of a toy marketed as the 'Superball' [17] whose unusual bounces are a consequence of strong translational-rotational coupling. Hefner [18] and Tavares [19] extended the treatment of Garwin's model to consider motion between two walls. The motivation of these authors appears to have been largely pedagogical although the model has been applied to phenomena in sports physics [20]. The third incarnation of the rough sphere occurred in 1993 when Broomhead and Gutkin [21] introduced the model to study the consequences of no-slip on the trajectories of a billiard on a confined geometry. Cox, Feres and Zhang [22] have extended the application of the rough sphere model in the study of mathematical billiards. Most recently, we demonstrated [23] that the rough sphere kinematics arises as a special case of the Cundall-Strack model [24] of a frictional particle widely used in the modelling of granular material.

The paper is organised as follows. In the next Section we shall derive the kinematic equations for the rough sphere between sliding walls. In Section 3 we shall consider the consequences of these equations for the case of stationary walls. In Section 4 we shall address the influence of the moving walls on the dynamics of the confined particle and in the final Section we shall present discussions and conclusions.

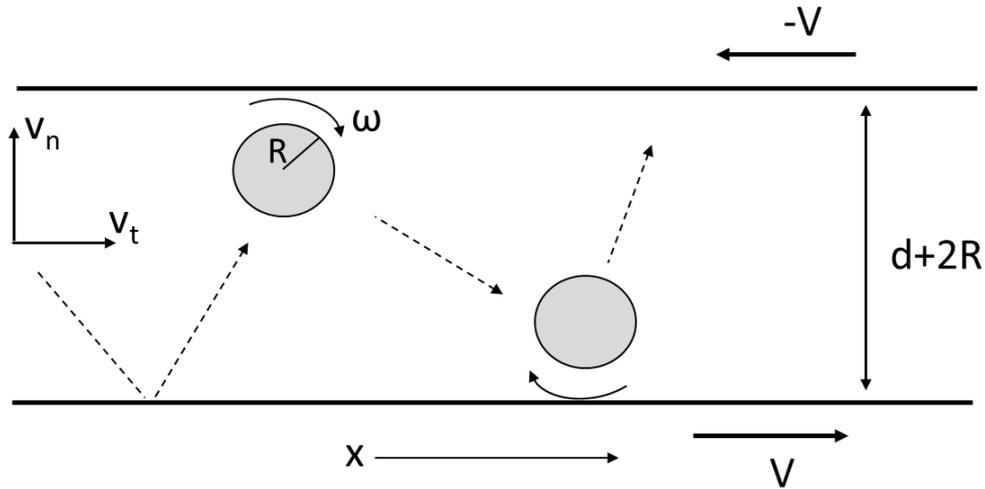

**Figure 1.** A sketch of the geometry of the particle-channel system indicating the width (d+2R) of the channel and the identification of the transverse and normal directions and the x coordinate. The bottom and top walls move along the x axes with velocities V and -V, respectively. Note that the contribution of the angular velocity to the velocity of the surface of the sphere relative to the wall changes sign between the top and bottom walls.

## 2. The Rough Sphere Kinematics

We shall consider a rough sphere of radius R and mass *m* moving between two parallel walls separated by a distance d+2R as indicated in Fig. 1. The translational velocity can be decomposed into components $v_n$ and $v_t$ along the normal and transverse directions, the latter we choose to lie along the x axes (see Fig. 1). The bottom wall moves with a velocity V along the x direction and the top wall moves with a velocity -V in the opposite direction.

Let $v_n(i)$ and $v_t(i)$ be the normal and transverse components of the translational velocity of the center of mass and $\omega(i)$ is the angular velocity , all after the *i*th collision. The first collision of the rough sphere and the wall is described by the kinematic expression

$$\begin{pmatrix} v_n(1) \\ v_t(1) \\ R\omega(1) \end{pmatrix} = \begin{pmatrix} -1 & 0 & 0 \\ 0 & M_{tt} & M_{tr} \\ 0 & M_{rt} & M_{rr} \end{pmatrix} \begin{pmatrix} v_n(0) \\ v_t(0) \\ R\omega(0) \end{pmatrix} \tag{2}$$

As the normal component of the translational velocity undergoes a simple sign reversal on collision and does not couple with the other degrees of freedom, it is not relevant for the analysis here. That leaves us with

$$\begin{pmatrix} v_t(1) \\ R\omega(1) \end{pmatrix} = M \begin{pmatrix} v_t(0) \\ R\omega(0) \end{pmatrix} \tag{3}$$

where the off-diagonal components of the matrix M represent explicit expression of the translation-rotation coupling due to the collision. The components of M have been obtained [16] by imposing the conservation of the kinetic and the angular momentum about the collision point to give

$$M = \frac{1}{1+\gamma} \begin{pmatrix} 1-\gamma & -2\gamma \\ -2 & -(1-\gamma) \end{pmatrix} \tag{4}$$

where $\gamma$ characterizes the mass distribution in the sphere through the expression for the moment of inertia, $I = \gamma m R^2$. The possible values of $\gamma$ range from zero up to 3/2, corresponding to the mass concentrated in the center or located in a thin shell at the sphere surface, respectively. For a sphere of uniform density, $\gamma = 2/5$. A note regarding reduced units - throughout this paper lengths are in units of $R$, mass in units of $m$ and time in units of $d/v_n$.

The second collision is with the opposite wall. As indicated in Fig.1, the rotational component of the transverse motion of the surface relative to the wall changes sign when we shift to reference frame of the opposite wall. What is top spin to the bottom wall is back spin at the top wall. This means that the second collision obeys the following equation,

$$\begin{pmatrix} v_t(2) \\ -R\omega(2) \end{pmatrix} = M \begin{pmatrix} v_t(1) \\ -R\omega(1) \end{pmatrix} \tag{5}$$

If, following Travares {19], we define a matrix

$$D = \begin{pmatrix} 1 & 0 \\ 0 & -1 \end{pmatrix} \tag{6}$$

then we have for the ith collision

$$\vec{v}_i = M\vec{v}_{i-1} \text{ for i odd} \tag{7a}$$

$$\vec{v}_i = D^{-1}MD\vec{v}_{i-1} \text{ for i even} \tag{7b}$$

where

$$D^{-1}MD = \frac{1}{1+\gamma}\begin{pmatrix} 1-\gamma & 2\gamma \\ 2 & -(1-\gamma) \end{pmatrix} \tag{8}$$

and $\vec{v}_i = \begin{pmatrix} v_t(i) \\ R\omega(i) \end{pmatrix}$.

The inclusion of the wall velocity V into the kinematics of the rough sphere collision is achieved by describing the collision in the reference frame of the moving surface. Within the moving wall reference frame, we impose energy conservation and the reversal of the particle surface velocity at contact, so that we have

$$\vec{v}_i = \vec{Q} + M(\vec{v}_{i-1} - \vec{Q}) \text{ for i odd} \tag{9a}$$

$$\vec{v}_i = -\vec{Q} + D^{-1}MD(\vec{v}_{i-1} + \vec{Q}) \text{ for i even} \tag{9b}$$

where $\vec{Q} = \begin{pmatrix} V \\ 0 \end{pmatrix}$ . Note that the change in reference state results in an adjustment of the effective transverse centre of mass velocity while leaving the angular velocity untouched, the latter point reflecting the separation of translational and rotational degrees of freedom. The sign associated with $\vec{Q}$ changes for the odd and even collision numbers because the two walls are moving in opposite directions.

## 3. Rough Sphere Dynamics Between Stationary Walls

We start our analysis with the case of stationary walls, i.e. V=0. This problem has been previously addressed by Travares [19] who demonstrated the surprising spatial confinement, earlier noted by Garwin [16], that arises from the translational-rotational coupling. The associated oscillations are conveniently captured with the map of the particle dynamics in the 2D phase space consisting of $v_t$ and $\sqrt{\gamma} R \omega$ . In Fig. 2 we present the phase space maps and trajectories in the physical space (x,y) for four different choices of $\gamma$, the mass distribution within the rough sphere . These phase points are kinematically connected at each collision as depicted in Fig.2.

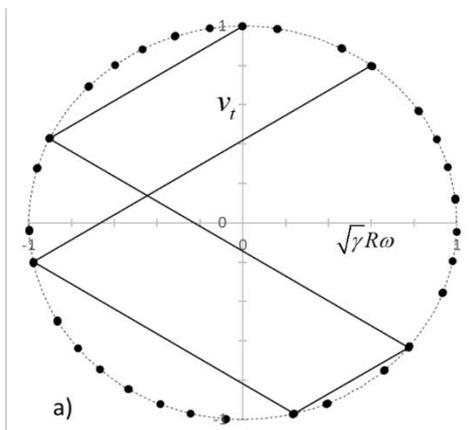

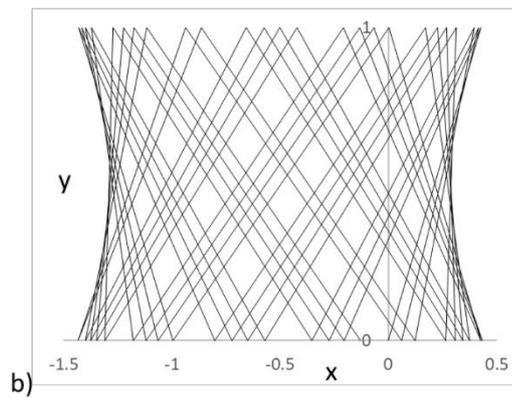

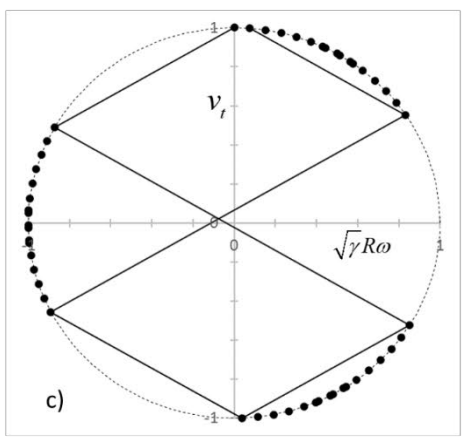

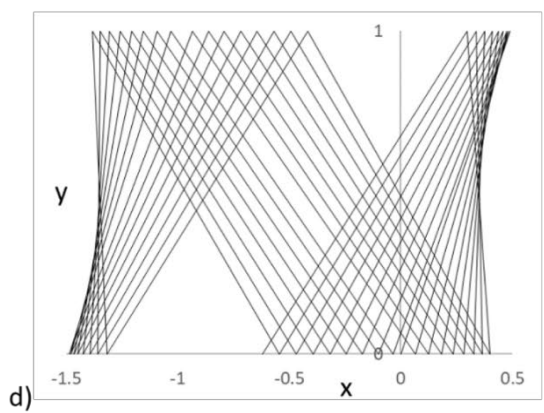

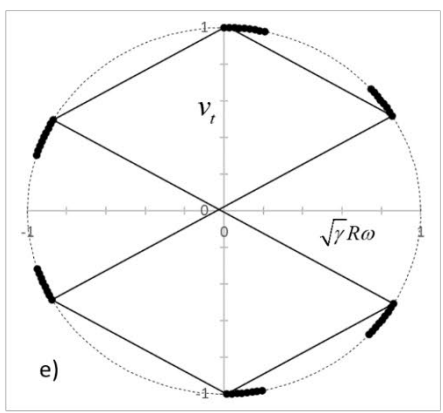

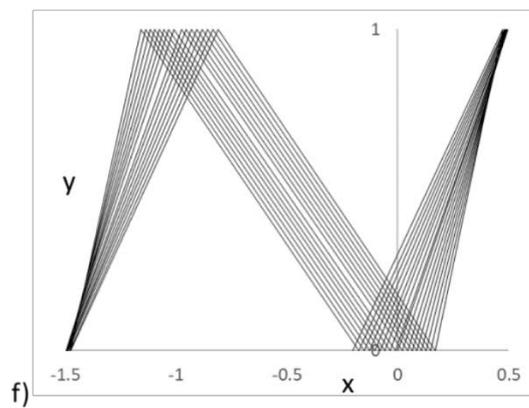

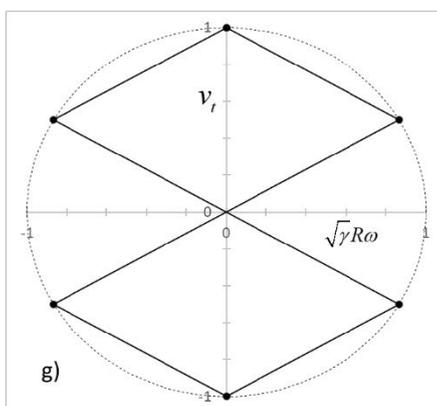

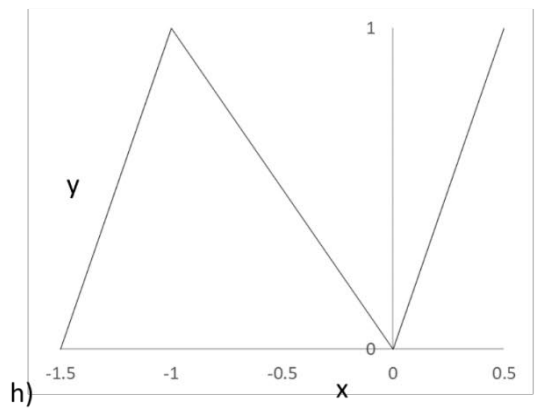

**Figure 2.** Maps of the phase space $\left(\sqrt{\gamma}R\omega, v_t\right)$ and trajectories of rough spheres for 4 values of γ: a) & b) 2/5 (uniform solid), c) & d) 0.3433, e) & f) 0.3367 and g) & h) 1/3. The links between subsequent phase space points are indicated as solid lines for a set of 6 consecutive collisions. In all cases the initial conditions were $v_t(0) = 1$ and $\omega(0) = 0$. In the trajectory maps, x = 0 corresponds to the position of the first collision with the (lower) wall.

We note the following features of the dynamics depicted in Fig. 2:

i) The large oscillations in the translational and rotational velocities are evident in Figs. 2a, 2e, 2c and 2g. The period τ, in number of collisions, is given by [19]

$$\tau = \frac{2\pi}{\cos^{-1}\left(\dfrac{1-\gamma}{1+\gamma}\right)} \tag{10}$$

which, for $2/5 \leq \gamma \leq 2/6$, ranges from 5.57 to exactly 6. A non-integer period results in an aperiodic oscillation as shown. Previously, Garwin [16] noted that the initial transverse velocity of a rough sphere between two walls is reversed roughly every 3 collisions, consistent with a period of roughly 6 collisions. Broomhead and Gutkin [21] pointed out that the bounded oscillations between parallel walls means that the trajectories of the rough sphere are stable with respect to perturbations, unlike the trajectories of the smooth elastic sphere. This means that the rough spheres trajectories in a fully enclosed area (a 'stadium') with parallel walls cannot be ergodic.

ii) The phase maps consist of a circle centred on the origin described by the energy conservation condition

$$v_t^2 + \gamma(R\omega)^2 = \frac{2}{m}E \qquad (11)$$

where E is the total kinetic energy of the particle. It follows that radii of the phase space circles in Fig. 2 are equal to $\sqrt{\frac{2E}{m}}$. As $\gamma$ approaches 1/3, a value associated with periodic motion, the phase space points concentrate about the 6 points of the periodic trajectory.

iii) The circular phase map provides an explicit picture of translation-rotation coupling. This coupling is imposed by the twin conditions – conservation of angular momentum and energy – in the instantaneous collision. The result, in the channel geometry, is to couple translations and rotations into a single coupled oscillation.

iv) The approach to the periodic orbit as $\gamma$ approaches 1/3 is marked by the continuous closing of the gap between the start and end of the 6-collision sequence.

v) While the average values of $v_t$ and $\omega$ are zero, both velocities fluctuate significantly about this mean, this is in spite of the particle being constrained to a non-slip boundary condition at the stationary walls. This observation underscores the point that the microscopic non-slip boundary applies to the detailed coupling of translational and rotational motion during the collision and does not represent a simple constraint on the translational velocity alone.

vi) The trajectories are bounded along the transverse direction. This trapping of the particle is a direct result of the roughly periodic reversal of the sign of the transverse velocity. The spatial extent of the bounded region is roughly independent of $\gamma$ down to a value of $\sim 0.1$, below which the bounded region steadily diverges, as shown in Fig.3a. The magnitude and position, relative to the point of first contact, of the bounded domain varies with the choice of the initial conditions (see Fig.3b).

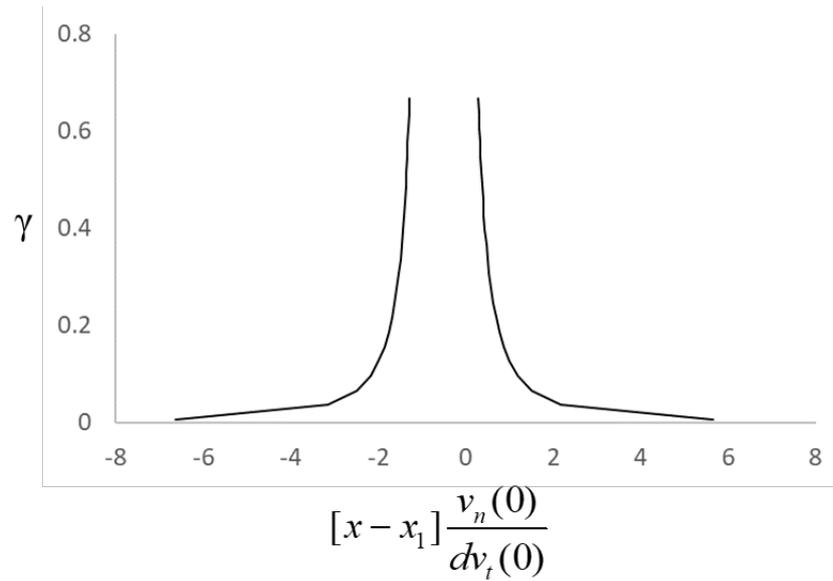

a)

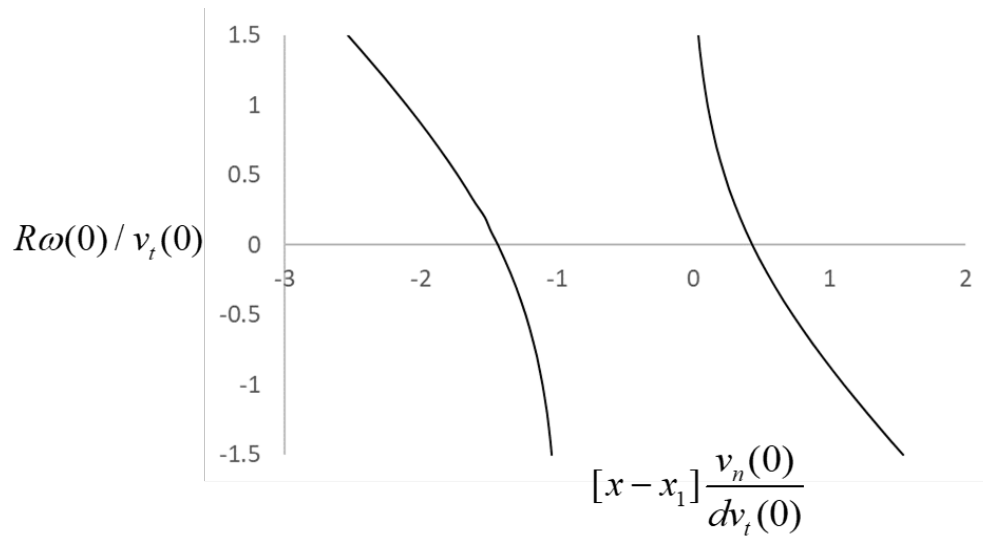

b)

**Figure 3.** a) The bounds along the x axes of particle motion as a function of $0 < \gamma \leq 2/3$ for the case where $\omega_0 = 0$ with x=0 marking the position of the first collision. b) The bounds of the motion along the x axes for the case of the solid sphere ($\gamma = 2/5$) for a range of values of $\frac{R\omega_0}{v_{t,0}}$.

The control of the position and width of the bound domain within the channel, demonstrated in Fig. 3b, is of particular significance in the case of a channel that has been truncated so that a sphere can be injected and ejected through this opening. If the end of the channel lies within the range of the particle's bounded motion, the particle will escape the channel as a rebound, as shown in Fig. 4. On the other hand, if the bounded region was completely contained within the truncated channel, the particle would be trapped. The result of this analysis (presented in the Appendix) is an outcome map (see Fig. 5) identifying those initial phase space values that result in a rebound and those that can result in trapping. (As discussed in the Appendix, the trapping here is marginal as, over time, the trajectories approach arbitrarily close to the open end of the channel.)

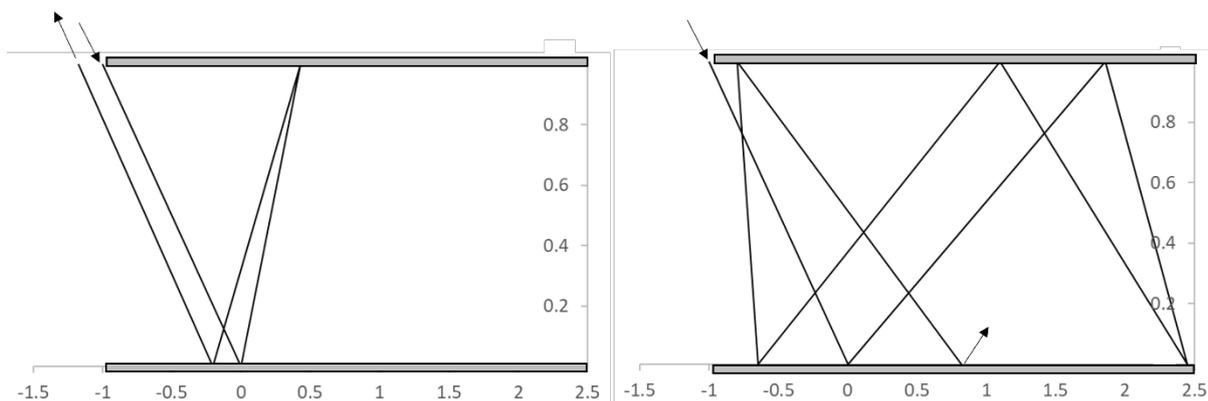

**Figure 4.** Examples of trajectories in the truncated channel: (left panel) a rebounding trajectory with initial conditions $[v_t(0) = 1, \omega(0) = 0]$, and (right panel) a trapped trajectory with initial conditions $[v_t(0) = 1, \omega(0) = -2.5]$ .

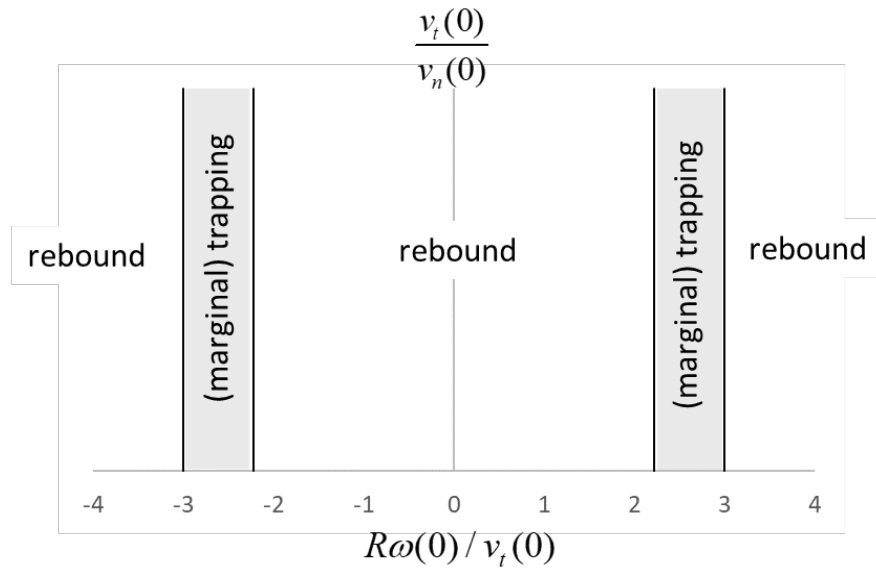

**Figure 5.** The outcome – rebound or trapping – of a rough sphere injected into the truncated channel as a function of the initial translational and rotational velocities. See Appendix for more details.

Given enough collisions, the non-periodic oscillations of the rough sphere between two walls will sample all points on the circular phase map (as shown in Fig. 6). The closer $\gamma$ is to a periodic value, the greater the number of collisions required.

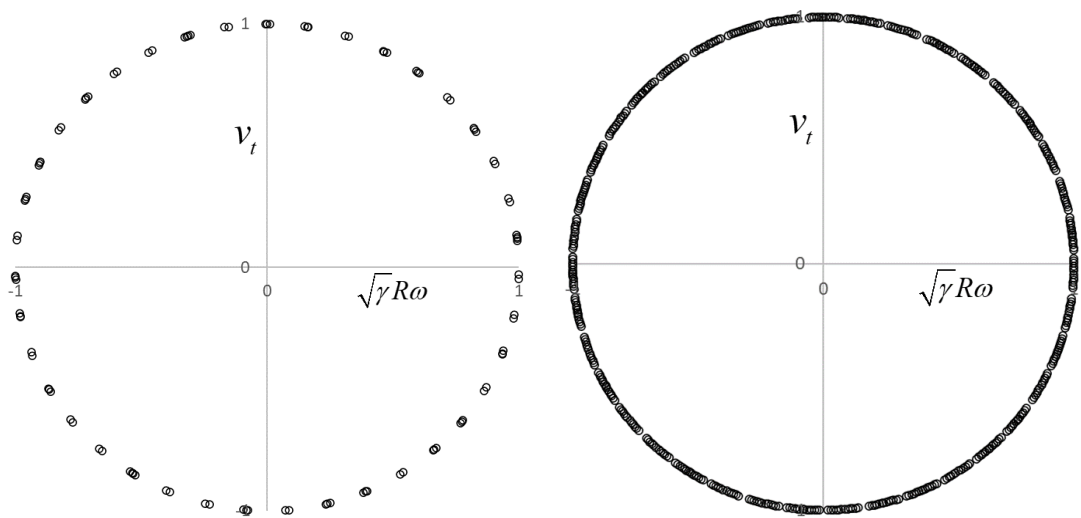

**Figure 6.** The coverage of the circular phase space as a function of the number of collisions for γ = 2/5 using 100 and 500 collisions, respectively.

A final observation regarding the oscillation of the trapped rough sphere. The reversal of the transverse velocity (and, hence, the particle trapping)  arises due to the switch in the sign of the rotational contribution to the particle surface velocity with each collision. The essential role played by rough collisions at both walls is clearly demonstrated if we treat the collisions with one wall as smooth, i.e. $v_t$ and $\omega$ are unchanged. As shown in Fig. 7, the trajectory of the particle in the rough + smooth channel  exhibits no reversal of the transverse velocity and, hence, no oscillations along the transverse direction. The influence of the single rough wall is simply to retard the propagation of the particle through the channel.

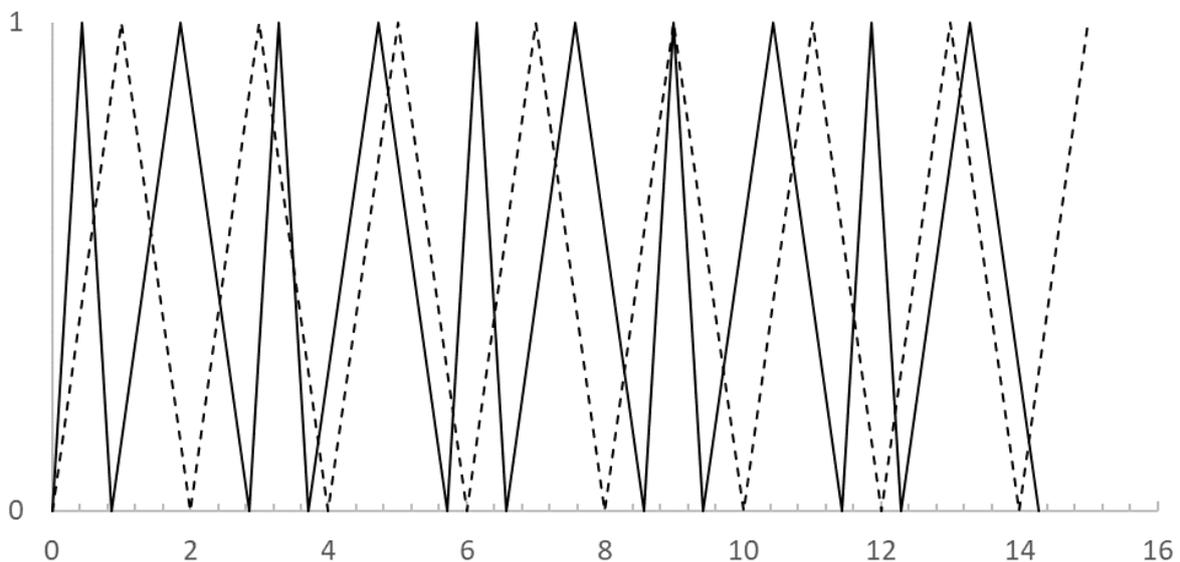

**Figure 7.** Plots of two trajectories, each starting with the same initial conditions ($v_t$ =1, ω =0) but with collisions characterised by a rough lower wall and a smooth upper wall  (solid line) or two smooth walls (dashed line).

## 4. Rough Sphere Dynamics Between Sliding Walls

Now we consider the case when the confining walls slide i.e. $V \neq 0$. In Fig. 8 we plot the phase space maps for V = 0, 1, 2 and 5 for the sphere with $\gamma = 2/5$. We find that the phase maps are still circles but now with a centre and radius that depends on V. The phase maps are described by the following equation,

$$v_t^2 + \gamma \left( R\omega - V \right)^2 = C_o \tag{12}$$

where $C_o = v_t^2(0) + \gamma(R\omega(0) - V)^2$. The associated particle trajectories take the form of oscillations bounded along the transverse direction, as in the case of the stationary walls, with the extent of the bound region increasing with increasing V.

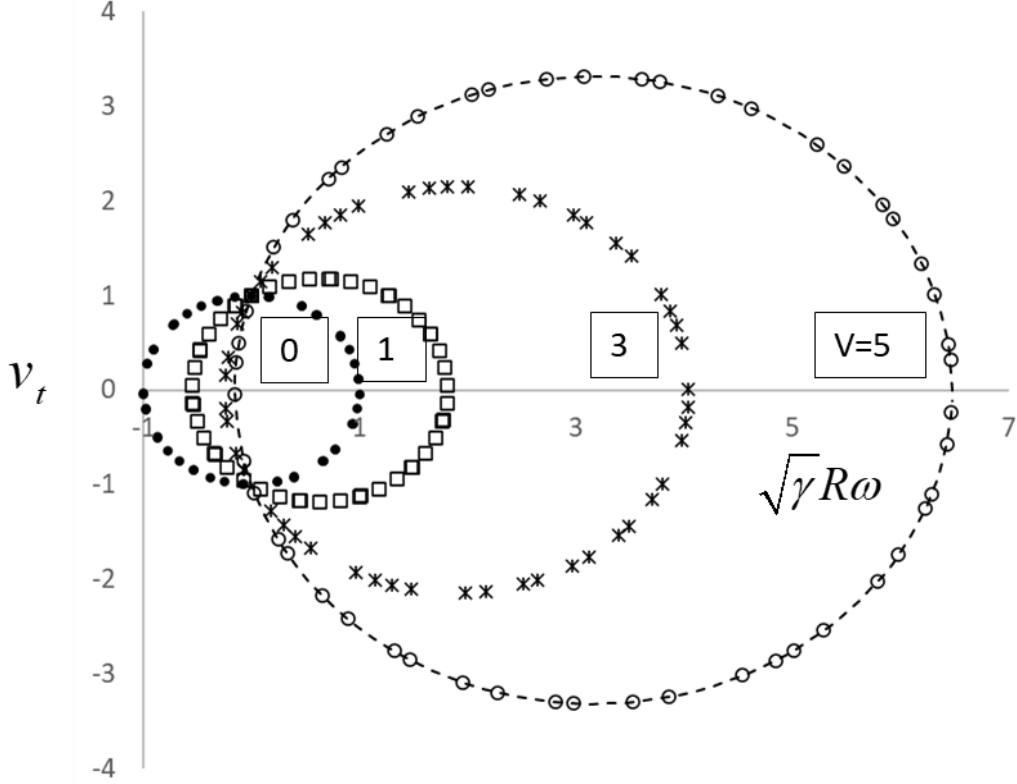

**Figure 8.** The phase space maps of the rough sphere between sliding walls with V = 0, 1, 3 and 5, as indicated. The dashed circle is calculated using Eq. 12 for V = 5. The rough sphere is a uniform solid (i.e. γ = 2/5) with $v_t(0) = 1$ and $\omega(0) = 0$.

Eq. 12, which represents our major result, is puzzling. The form of the equation indicates that the quantity $C_o$ is conserved but no such conservation was ever imposed. Furthermore, the V-dependent shift in the mean rotational velocity is surprising given that the reference frame shift introduced in Eq. 9 was restricted to the translational velocity. We shall now address these issues.

It is useful to start by establishing the existence of a fixed point in the phase space – the values of the initial velocities that remains unchanged by collision. The fixed point for our confined rough sphere is

$$\begin{pmatrix} v_t(0) \\ R\omega(0) \end{pmatrix} = \begin{pmatrix} 0 \\ V \end{pmatrix},$$

(13)

a result that can be verified by substitution into the kinematic equations Eq. 9 for both odd and even collision numbers. The assignment of the value V to the rotational velocity in Eq. 13 is necessary to offset the similar term linked to $v_t$ in Eq. 9. Eq. 13 is fixed point for both odd and even collisions because rotational velocity and wall velocity undergo the same sign flip as the particle moves between top and bottom walls.

Having established the existence of a phase space fixed point (i.e. the center of the phase circle) we can then write an arbitrary set of velocities as

$$\begin{pmatrix} v_t \\ R\omega \end{pmatrix} = \begin{pmatrix} 0 \\ V \end{pmatrix} + \begin{pmatrix} v_t \\ R\omega - V \end{pmatrix} \tag{14}$$

which, when substituted into the kinematic equations, results in

$$\begin{pmatrix} v_{t,i} \\ R\omega_i - V \end{pmatrix} = M \begin{pmatrix} v_{t,i-1} \\ R\omega_{i-1} - V \end{pmatrix} \text{ for i odd} \tag{15a}$$

$$\begin{pmatrix} v_{t,i} \\ R\omega_i - V \end{pmatrix} = D^{-1}MD \begin{pmatrix} v_{t,i-1} \\ R\omega_{i-1} - V \end{pmatrix} \text{ for i even} \tag{15b}$$

which are just the equations for the system with stationary walls, the reference frame shifts arising from the wall movement have been accounted for by the fixed point component and so have disappeared from Eq. 15. Since, by construction, kinematic equations of the form of Eqs. 15 conserve the associated energy, Eq. 12 follows directly.

The total energy, we emphasise, is *not* conserved when the walls move. This is immediately evident in the plot of energy as a function of collision number for various values of V. As shown in Fig. 9a, the total energy is not a constant of the motion but oscillates when $V \neq 0$ with a principal period of $\sim 3$ collisions (the period for the reversal of the transverse velocity, already identified) and longer frequency of $\sim 25$ collisions. This slower frequency process corresponds to the period for the sampling of the maximum (or minimum) rotational velocity.

The mean total energy E, as plotted in Fig. 9b, increases quadratically with V according to Eq. 12, i.e.

$$\frac{2}{m}E = <v_t^2> + \gamma < (R\omega)^2 > = 2\gamma V^2 - 2\gamma VR\omega(0) + \frac{2}{m}E_o \qquad (16)$$

where $E_o$ is the initial total energy. This increase is dominated by the energy deposited specifically into the rotational degree of freedom. This dramatic breakdown of the equipartition of the energy between the modes is clearly demonstrated in Fig. 9b. Based on Eq. 12, we can write down an explicit relationship between the average rotational and translational energies,

$$\gamma \left\langle (R\omega)^2 \right\rangle = \gamma V^2 + \left\langle v_t^2 \right\rangle \qquad (17)$$

This excess rotational energy arises directly from the shift in the phase space fixed point described in Eq. 13. The rotation is specifically selected by the applied shear because they share a common symmetry when the shearing system is viewed globally. In contrast to the localized constraint that is described by the macroscopic boundary conditions, the microscopic implementation of non-slip has coupled the rough sphere degrees of freedom to the global arrangement of the driven walls. The result is a particle forced to spin as a result of its alternating interaction with the two walls.

How do results compare with the expectation of flow bounded by macroscopic stick boundaries? As these flows are typically expressed in terms of the translational velocities, we shall focus on this component. If we assume a flow field $v_t(y) = -2Vy$ along the direction normal to the walls, then $<v_t^2> \propto V^2$. The analogous value for the rough sphere systems is

$$<v_t^2> \propto \left( v_t^2(0) + \gamma(R\omega(0) - V)^2 \right) \qquad (18)$$

While both expressions describe a quadratic dependence of the translational kinetic energy on V, there are some significant differences arising from the translation-rotation coupling. For the rough sphere, $<v_t^2>$ becomes independent of V when $R\omega(0) = V$, highlighting the fact that microscopic energy transfer from the wall to the adjacent particle is controlled by the rate at which the particle surface is moving relative to the wall.

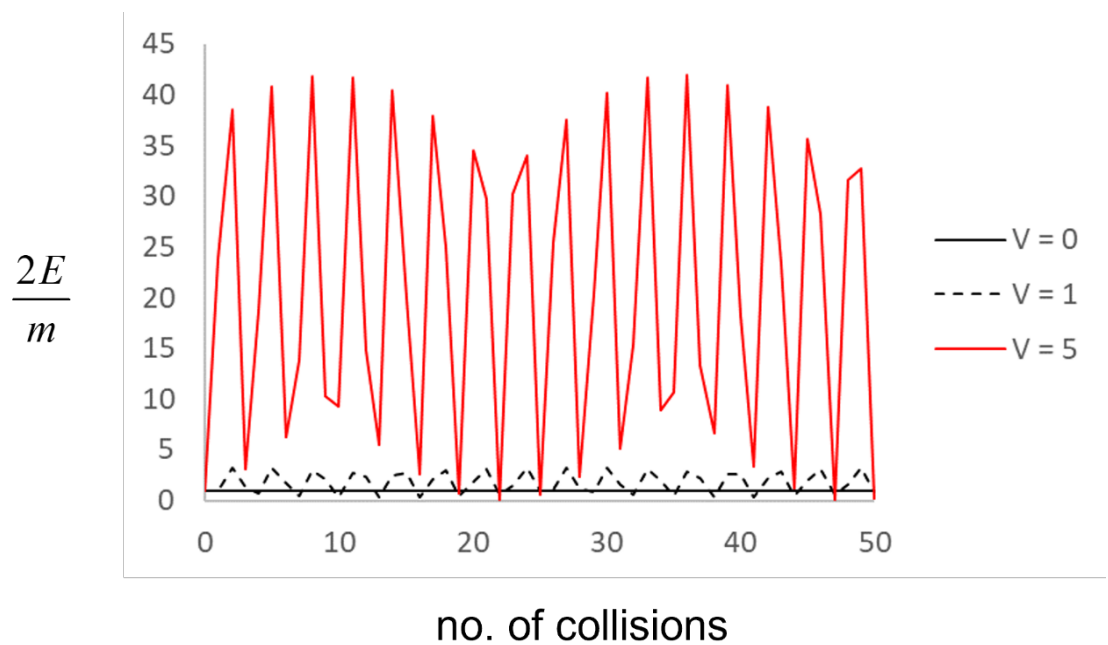

a)

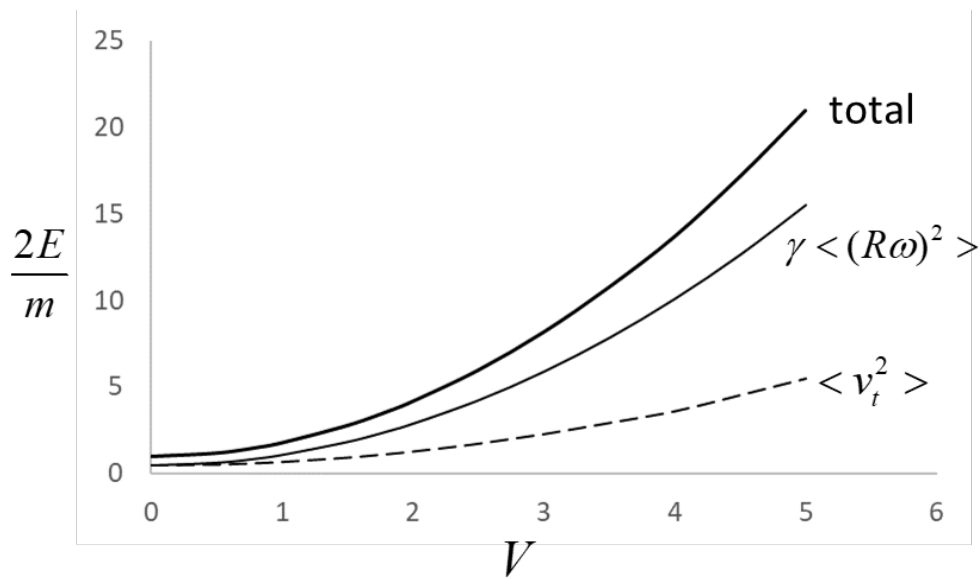

b)

**Figure 9.** a) The total energy as a function of collision number for V = 0, 1 and 5. b) The mean total, translational and rotational energy as a function of V. In all cases the initial conditions are $v_t(0) = 1$ and $\omega(0) = 0$.

## 5. Conclusions

As described in the Introduction, the implementation of realistic physical interactions between particle and the confining wall typically leads to more complex boundary conditions (e.g. shear rate dependent slip, etc) than the simple stick boundary typically employed in modelling macroscopic hydrodynamic flows. In this paper, we have examined a model case in which a non-slip boundary condition can be directly applied at the microscopic level. Despite the explicit imposition of stick conditions at the level of individual collisions, we still find more complex behavior than that predicted by the conventional stick boundary condition. In the case of the rough sphere, this complexity stems directly from the coupling of translation and rotation that results from the non-slip kinematics, a coupling that effectively binds $v_t$ and $\omega$ together into a single oscillatory mode. The periodic reversal of the translational velocity results in the particle being trapped within an interval along the transverse direction. An observable consequence of this transverse oscillation is the rebound of a particle injected into the open end of the channel. We have established that trapping remains a possibility for the open channel but only for a narrow range of initial rotational velocities.

Sliding the walls past one another does not qualitatively alter this picture of bounded oscillations. The consequences of moving the walls past one another with a relative velocity of 2V is to increase the average rotational velocity to $<\omega> = V / R$ and to alter the oscillatory component of the energy $C_o$ (see Eq. 12). This specific selection of rotations over translations for energy deposition arises due to the rotational character of the shearing motion of the two walls that permits direct coupling with the particle rotation. It is likely that, as the density increases, the rough sphere rotations will become increasingly coupled to dynamics of surrounding particles and less susceptible to the influence of the sliding walls. How this crossover in behaviour takes place with increasing density, particularly its connection with theories [25] that include rotational velocities at the coarse-grained level of hydrodynamics, is an interesting question for further study.

 A final observation regarding the extension of these results to particles interacting via potentials with non-zero length scales. As discussed in ref. [23], the mapping of molecular dynamics to the rough sphere kinematics is delicate. Two conditions need to be met. The first is that the particle-wall interaction must generate some sort of transverse force/torque. Typically, this results in a transient roto-vibrational oscillation. The second condition is that the collision duration must be equal to the period of this transverse oscillation. If not, energy will be dissipated in the collision. This paper was motivated by the need to set aside many of the complexities of collisional coupling between rotations and vibrations. Our results, particularly the strong coupling between the sliding walls and the particle rotation, suggest that more realistic models of molecule-wall collisions may result in a surprising variety of dynamic consequences.

**Acknowledgements**

YW gratefully acknowledges support in the form of an Australian Postgraduate Award from the Australian Federal Government.

**Appendix: Particle Rebound from a Truncated Channel**

We have not addressed how the particle motion in the channel is initiated. An *in situ* injection of a particle into the infinite channel is conceivable e.g. by desorption of a particle from one of the walls. An alternative would be an injection of the particle through an open end of the channel. This requires that we truncate the channel and sets an upper limit on how far $x_1$, the position of the first particle-wall collision, can be from this opening. As shown in Fig. 10, the maximum distance of the point of first collision $x_1$ is $d\dfrac{v_t}{v_n}$ where the separation between the walls is d+2R. From the analysis of Section 3, we can establish the transverse bounds of the particle trajectory relative to the point of initial collision. If the distance between the opening and $x_1$ lies within this bound then the particle will escape from the channel through the same opening by which it entered. The result is a rebound out of the channel. In the case of initial conditions for which the bound motion does not extend as far as the opening, the particle enters the channel and then is permanently trapped.

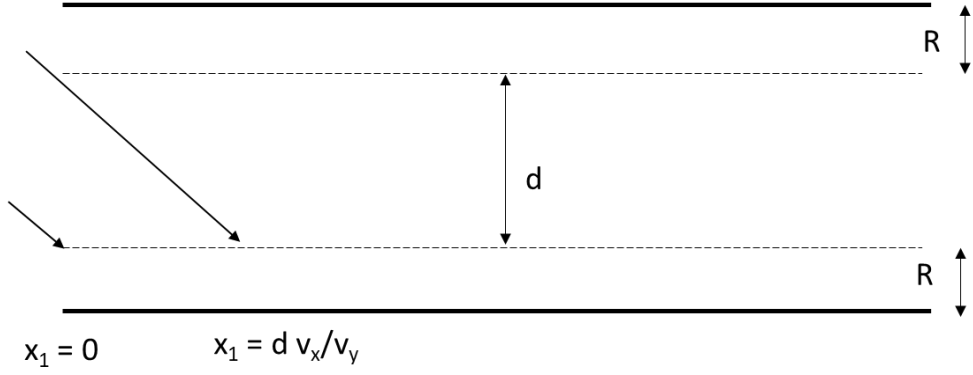

**Figure 10.** A sketch showing the geometrical definition of the maximum and minimum values of $x_1$, the position of the first collision with the channel wall, of particle entering from an open end.

Travares [19] has derived an analytic express for $x_n$ of the particle along the transverse axes at the nth collision,

$$\Delta_n = x_n - x_1 = \frac{A\Delta t}{2}\left[\frac{\sin((n+1)\theta/2)\cos(\phi+n\theta/2)}{\sin(\theta/2)} - \cos(\phi) - \cos(n\theta+\phi)\right] \quad \text{(A1)}$$

where

$$A = \sqrt{v_{0,t}^2 + \gamma\overline{\omega}_0^2}$$
$$\phi = \arctan\left(\sqrt{\gamma}\,\overline{\omega}_0\,/\,v_{0,t}\right) \quad \text{(A2)}$$

$$\cos(\theta) = \frac{1-\lambda}{1+\lambda} \quad \text{(A3)}$$

If we define $P = \sqrt{\gamma}\,\dfrac{R\omega(0)}{v_t(0)}$ and $M = \dfrac{v_t(0)}{v_n(0)}$, then we can write

$$\Delta_n = d\,\frac{M}{2}\sqrt{1+P^2}\,f(n,P)$$
$$x_1 = dM \quad \text{(A4)}$$

A particle, injected as deeply as possible into the channel, will rebound if $\frac{\Delta_n}{x_1} < -1$ for any

value of n >1. From Eq. A4 we have

$$\frac{\Delta_n}{x_1} = \frac{1}{2}\sqrt{1+P^2}\,f(n,P) \qquad\qquad\qquad (A5)$$

which establishes that the condition for rebound is independent of M and depends only on P.

We have calculated the minimum value of $\frac{\Delta_n}{x_1}$ for a range of values of P and plotted the

results in Fig. 11. We find that the condition for rebound is satisfies over all of the range of P

except for the interval -3.0 < P < -2.2 where the bound of the particle motion lies extremely

close to the end of the channel. In this limited range of initial angular velocities, the

possibility exists for the particle to remain trapped in the channel, at least for a significant

number for collisions before escaping. We call this situation *marginal trapping*. Note that is

result is for the case that the first collision occurs at the lower wall. For a particle colliding

with the upper wall, we have to flip the sign of P so that the marginal domain for an initial

collision on the upper wall is 2.2 < P < 3. If we combine both types of collisions, we

conclude that marginal trapping is possible for 2.2 <|P| < 3 as shown in Fig. 5. This means

that the regions denoted '(marginal) trapping' in Fig. 5 will include both marginal trapping

and rebound trajectories depending on whether the particle first collided with the lower or

upper wall, respectively. For all other choices of initial conditions, the rough sphere will be

ejected from the truncated channel by the same opening through which it entered

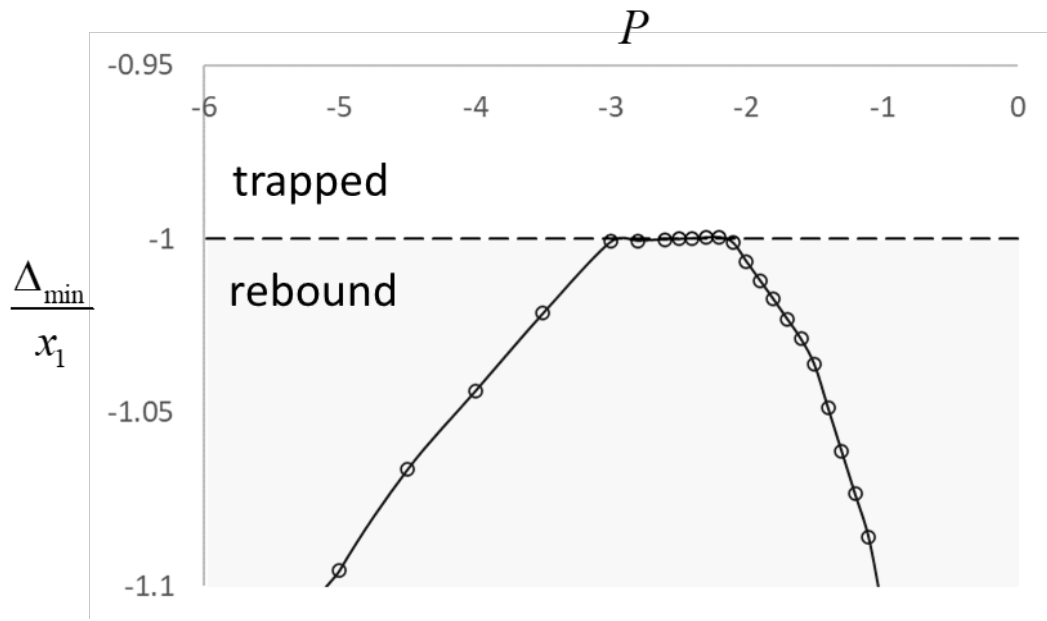

**Figure 11.** The minimum value of $\dfrac{\Delta_n}{x_1}$ plotted as a function of $P = \dfrac{R\omega(0)}{v_t(0)}$. As explained in

the text, the threshold between rebound or trapping, as indicated by the dashed line, is

$$\frac{\Delta_n}{x_1} = -1$$